\documentclass[conference]{IEEEtran}
   \usepackage{mathptm}
   \usepackage{caption}
   \usepackage{epsfig}
   \usepackage{subfigure}
   \usepackage{color}
   \usepackage{latexsym}
   \usepackage{pslatex}
   \usepackage{algorithm}
   \usepackage{algorithmic}
   \usepackage{multirow}
   \usepackage{balance}

\begin{document}

\title{Logic BIST: State-of-the-Art and Open Problems}

\author{
\IEEEauthorblockN{Nan Li}
\IEEEauthorblockA{
School of ICT \\
Royal Institute of Technology \\
164 40 Stockholm, Sweden \\
nan3@kth.se
}
\and
\IEEEauthorblockN{Gunnar Carlsson}
\IEEEauthorblockA{
Development Unit Radio \\
Ericsson AB \\
164 80 Stockholm, Sweden \\
gunnar.carlsson@ericsson.com
}
\and
\IEEEauthorblockN{Elena Dubrova}
\IEEEauthorblockA{
School of ICT \\
Royal Institute of Technology \\
164 40 Stockholm, Sweden \\
dubrova@kth.se
}
\and
\IEEEauthorblockN{Kim Peters\'en}
\IEEEauthorblockA{
Development Unit Radio \\
Ericsson AB \\
164 80 Stockholm, Sweden \\
kim.petersen@ericsson.com
}
}

\maketitle

\begin{abstract}
Many believe that in-field hardware faults are too rare in practice to justify the need for Logic Built-In Self-Test (LBIST) in a design. Until now, LBIST was primarily used in safety-critical applications. However, this may change soon. First, even if costly methods like burn-in are applied, it is no longer possible to get rid of all latent defects in devices at leading-edge technology. Second, demands for high reliability spread to consumer electronics as smartphones replace our wallets and IDs. 
However, today many ASIC vendors are reluctant to use LBIST.
In this paper, we describe the needs for successful deployment of LBIST in the industrial practice and discuss how these needs can be addressed. Our work is hoped to attract a wider attention to this important research topic.  
\end{abstract}


\section{Introduction}  

A wide-spread opinion is that in-field hardware faults are too rare in practice to justify the need for adding Logic Built-In Self-Test (LBIST) into a design.
Until now, LBIST found its use mainly in safety-critical (automotive, medical, military),
mission-critical (deep-space, aviation) and high-availability (telecom) applications.
However, this may change soon. We expect that, in process technologies below 22nm, LBIST will become 
compulsory for Application-Specific Integrated Circuits (ASICs), Application-Specific Standard Products (ASSPs) and complex commercial ICs. The reasons for this are twofold.

On one hand, even if costly methods like burn-in are applied, 
it is no longer possible to get rid of all latent defects in highly complex devices at leading-edge technology nodes~\cite{ITRS}. A latent defect which is not visible during production testing may lead to a fault later on during the lifetime of the device. 

On the other hand, we are entering a new era of technology where electronic devices take control of many aspects of our lives. As smartphones replace our wallets and IDs, their high reliability and security become a must. LBIST can be used to quickly detect a fault and trigger self-repair in order to maintain reliability and security at a high level. 

Apart from detecting faults in-field, LBIST helps reduce rising costs of production testing by complementing the traditional external testing methods.
Today test cost can be more than one third of the product cost~\cite{HeF99}, and it is likely to grow in the future. 
On one hand, the number of transistors per chip pin increases which each generation of process technology~\cite{Mo65}. On the other hand, technologies at 45nm and below are prone to small delay defects and therefore require more tests~\cite{YiTC11}. As a consequence, the amount of test data per chip pin is growing. For consumer electronics, the test data volume in 2018 is expected to be 6 times larger than the one in 2013~\cite{ITRS}. On the contrary, the minimum size of the Automatic Test Equipment (ATE) memory is expected to grow only by 33\%~\cite{ITRS}. Until now, test compression~\cite{RaTKM04} has effectively combated the increase in test data volume by exploiting don't care values in ATPG-generated test patterns for digital logic. However, compression is fundamentally limited by the number of specified bits in the test patterns~\cite{BaT07,Wi08}. The challenges are rising to a new level with roll-out of 3D devices with very little physical access. 

Despite of numerous advantages of LBIST, today many ASIC vendors are reluctant to use it due to problems with non-relevant faults caused by high levels of switching activity
in test mode, as well as difficulty to reach a sufficient test coverage for some designs.
Furthermore, although existing LBIST techniques are good at detecting random faults, they
do not provide adequate protection against malicious circuit alternations known as hardware Trojans.
For example, a recent attack on Intel's Ivy Bridge processor demonstrated that the traditional LBIST
may fail even the simple case of stuck-at fault type of Trojans.

In this paper, we describe the needs for successful deployment of LBIST in the industrial practice
and discuss how these needs can be addressed. Our goal is to attract a wider attention to this important research topic.  

The paper is organized as follows.  
Section~\ref{adv} summarises advantages of LBIST.
Section~\ref{dis} describes disadvantages of LBIST.
Section~\ref{prev} presents the previous work. 
In Section~\ref{needs} we discuss what needs to be done to overcome problems associated with LBIST and give our thoughts on how these problems can be addressed.
Section~\ref{con} concludes the paper.

\section{Background}

This section gives a brief introduction to testing and LBIST. It is intended for readers not familiar with testing. For more details, please see~\cite{AbBF94,RaT98}. 

Testing has been an important topic in electronic industry since the first transistor was created. Testing is used as a part of design validation, as a quality indicator for manufacturing process control, and for the detection of defective chips prior shipping them to a customer. Taking into account that the manufacturing yield  can be as low as 30\%, testing chips for defects at the manufacturing stage is unavoidable.

Production testing is performed by applying a test pattern to a circuit under test and comparing the resulting response with the expected correct response. The test execution time consists of
three components: time to generate test patterns, time to apply them, and time to compute the
circuit response.

Ideally, production test cost should take a negligible part of the overall developing and manufacturing cost. In the best-case scenario:
\begin{enumerate}
\item Test development and execution should be fully automated and take essentially no time.
\item Testing equipment should be very inexpensive.
\item Test coverage should be 100\%.
\end{enumerate}

However, in reality, production test cost can be more than one third of the overall cost of a chip today and this number is likely to grow in the future~\cite{HeF99}. The reason for this is that general-purpose ATE is expensive and slow. Thus, as test data volume increases, test application time grows prohibitively long.
Test application time is directly related to the test cost, e.g. 3-5 cents can be charged per second of ATE time. Moreover, if test data volume is too large to fit the internal ATE memory, then
the use of more advanced ATE is required. This further increases test cost.

\begin{figure}[t]
\begin{center}
\includegraphics*[width=2.4in]{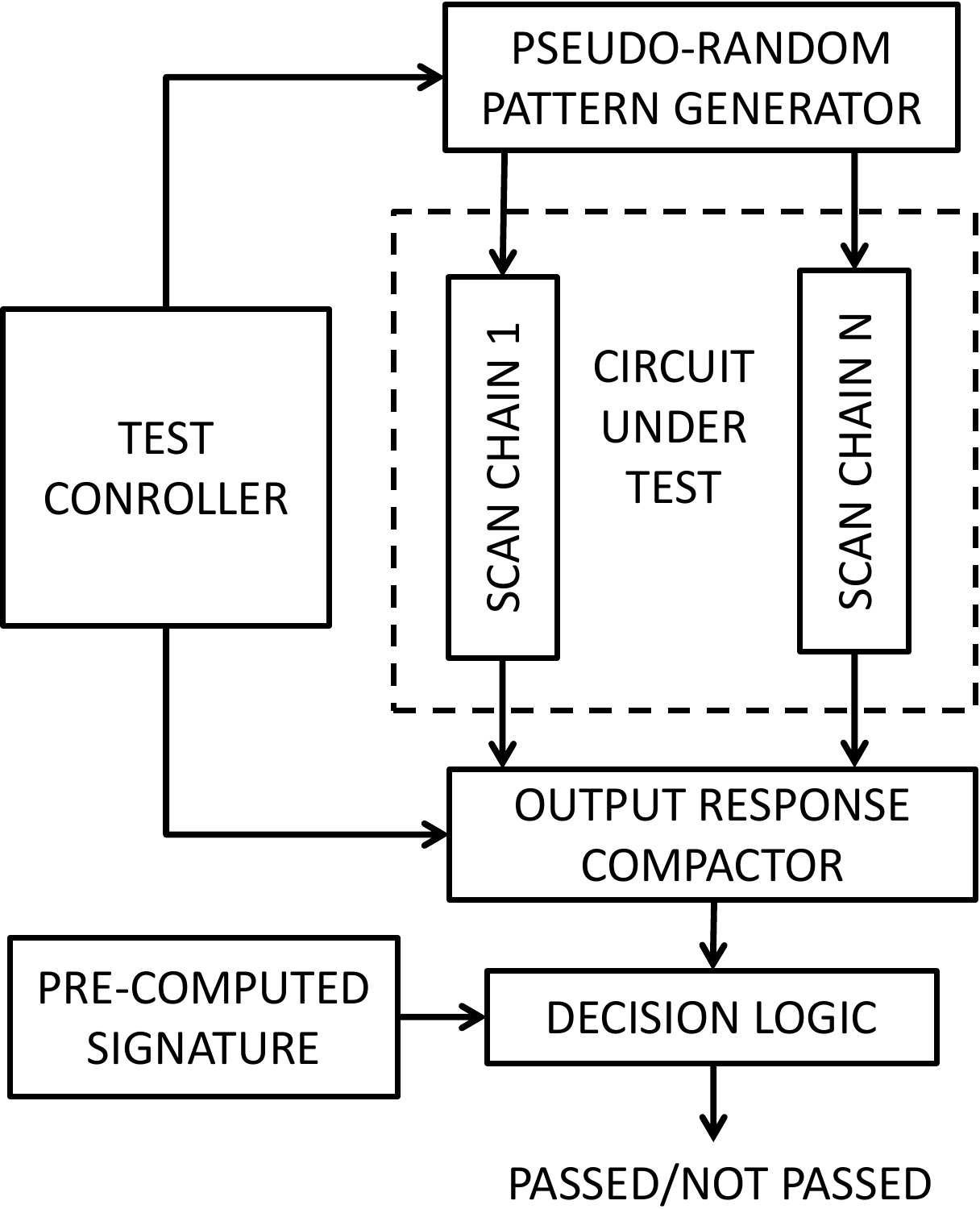}
\caption{The structure of traditional LBIST.}\label{lbist}
\end{center}
\end{figure}

Built-In-Self-Test (BIST) attempts to reduce the raising complexity of external testing by 
incorporating test generation and response capture logic on-chip~\cite{McC85}. On-chip circuitry usually works at a much higher frequency than
a tester. So, by embedding the test pattern generator on chip, we can reduce test application time. In
addition, by embedding the output response analyzer on chip, we can reduce time to compute the circuit
response.

There are different types of BIST. Logic BIST (LBIST), on which we focus in this paper, is used for testing random digital logic~\cite{RaT98}. Memory BIST (MBIST) is designed for testing memories~\cite{Sh02}.

LBIST typically employs a Pseudo-Random Pattern Generator (PRPG) to generate test patterns that are applied to the circuit's internal scan chains and an output response compactor for obtaining the compacted response of the circuit to these test patterns, called {\em signature} (see Figure~\ref{lbist}). Faults are detected by comparing the computed signature to the expected ``good'' signature.
 
In theory, it is possible to generate a complete set of test patterns off-line using some Automatic Test
Pattern Generation (ATPG) method and store this test set in an on-chip Read Only Memory (ROM). However, such a
scheme does not reduce the cost of test pattern generation and requires a very large ROM. Several gigabits of
test data may be required for a multi-million gate design~\cite{HeF99}.

\begin{figure}[t]
\begin{center}
\includegraphics*[width=3.5in]{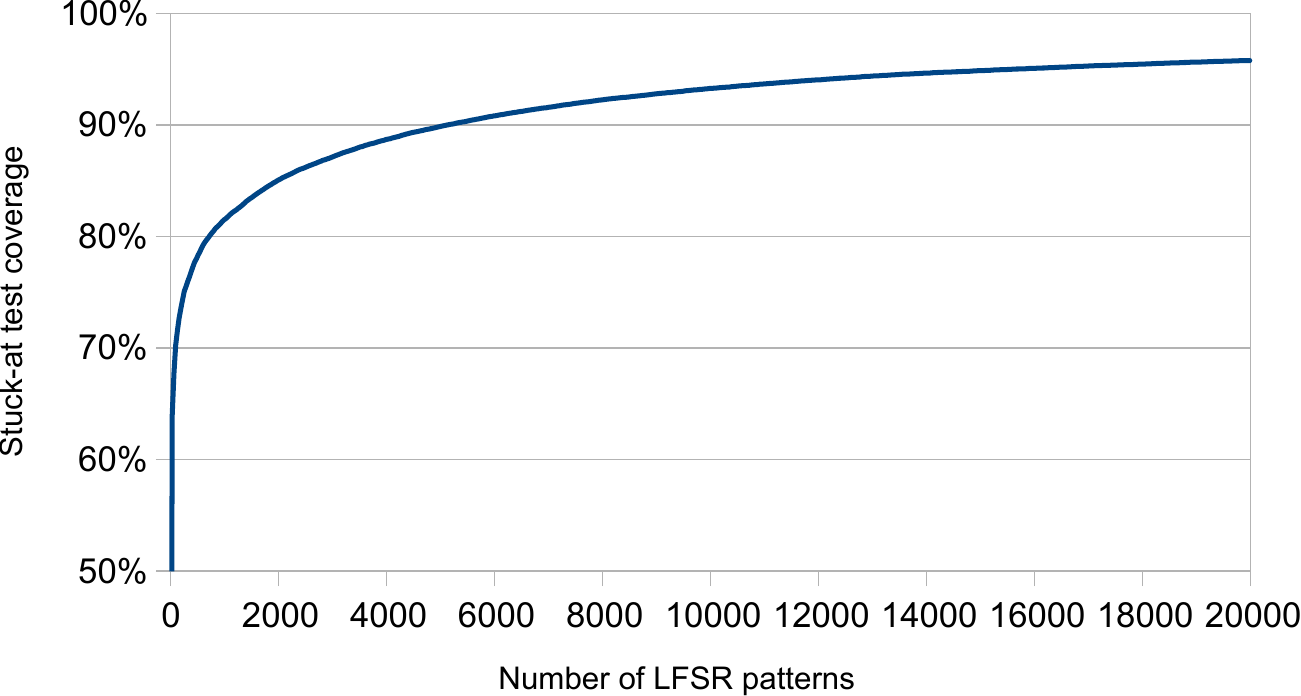}
\caption{Stuck-at test coverage of Leon3 processor as a function of the number of LFSR patterns.}\label{leon_cov}
\end{center}
\end{figure}

Instead, pseudo-random patterns generated by
an LFSR are usually used as test patterns~\cite{McMMW88}. 
LFSRs are simple, fast, and easy to implement in hardware~\cite{Golomb_book}.
Therefore, the area overhead of an LFSR-based LBIST is
low. However, its test execution time may be long. As we mentioned above, the test execution time depends on number of test patterns which need to be applied to reach a satisfactory fault coverage. There is a class of faults called {\em random-pattern resistant} faults which are hard to detect with pseudo-random patterns generated by an LFSR~\cite{Da98}. 
Typically, test coverage grows logarithmically with the number of LFSR patterns: 
first it increases quickly, then it flattens, and finally it reaches its saturation. 
Figure~\ref{leon_cov} shows the stuck-at test coverage of Leon3 processor with 8 cores as an example. 
We can see that a saturation point near 96\% is reached after the application of 20.000 LFSR patterns.
In Section~\ref{prev} we describe techniques which has been proposed for increasing test coverage of LBIST.

The output response compactor is usually implemented by a  
Multiple Input Signature Register  (MISR). Since the output response is compacted, it is possible that a faulty circuit may produce the same signature as the correct circuit. This is known as an {\em aliasing} error. If an MISR with a primitive generator polynomial is used\footnote{An irreducible polynomial of degree $n$ is called primitive if the smallest $m$ for which it divides $x^m+1$ is equal to $2^n-1$\cite{Golomb_book}.} then the aliasing probability is bounded by $1/2^n$~\cite{DaOFE90}, where $n$ is the length of the MISR.

\section{Advantages of LBIST} \label{adv}

LBIST requires no interaction with a large, expensive ATE. Only a small low-cost tester is needed to start the testing. Furthermore, LBIST can be re-used at many levels: board, subsystem and system. It can take an advantage of identical cores/blocks on a chip. 

LBIST is also re-usable at different stages during system life-cycle, including in-field.
For example, many cars today contain several electronic units that control engine, brakes, steering, airbags, etc. Every time a car is turned on, all units are self-tested~\cite{MePZ12}. 
If implemented properly, LBIST can be very helpful in locating the smallest defective unit in a system, which is a great advantage for today's complex systems.

In LBIST test patterns are applied at-speed, which is crucial for detecting timing faults in 65nm technologies and below. In the past, at-speed testing was used primarily for very high performance designs, e.g. full custom microprocessors. However, higher clock frequencies combined with process scaling made at-speed testing a necessity for common designs with ASICs~\cite{Ma13}.

In high-availability applications such as telecom, LBIST helps reducing downtime. 
The failure of a board with LBIST can be quickly diagnosed in-field by taking it out of the operation and running self-tests. If self-tests detect a fault, the board can be replaced by a spare and the system re-started. This reduces Mean Time to Repair (MTTR) and, hence, increases 
the availability of a system, which is given by
\[
A(\infty) = \frac{MTTF}{MTTF+MTTR}
\]
where MTTF is Mean Time to Failure.

Finally, LBIST enables test-ready Intellectual Property (IP) since the test data can be built into the chip from the beginning. 


\section{Disadvantages of LBIST} \label{dis}

However, nothing comes without a cost. LBIST also has a number of drawbacks.

First, LBIST logic increases chip area. If an LFSR is used for generating test patterns, the LBIST area overhead is typically 1\% of chip area or less. Given that other design-for-test techniques cause a significantly larger area overhead, e.g. about 10\% for scan and at least 20\% for MBIST, 1\% area overhead of LBIST is typically acceptable.

Second, LBIST requires that the design is either cleaned from unknown states that might violate the signature (X-bounding), or that such states are blocked using an extra circuitry. 
The designer who is inserting X-bounds should be aware of the sources of Xs in a design (e.g. floating ports of LBIST partitions, non-scan flops' outputs, latch outputs, non-scan hard blocks outputs, analog blocks outputs). A methodology to tackle this problem exists~\cite{Xbound}, so, again, X-bounding is not the real issue with LBIST.
 
The real issue with LBIST is that pseudo-random pattern-based testing 
toggles much more logic compared to functional operation, resulting is excess power dissipation.
During LBIST, the entire circuit is usually configured into a scan mode and test vectors are shifted into the scan chains ({\em shift cycle}) and then applied to simulate the functional operation ({\em capture cycle}). The loading and unloading of scan chains during shift cycle increase switching activity. Capture cycle also typically has a higher switching activity compared to the one of the functional operation, e.g.~\cite{Sa03} reports a two-fold increase in peak power dissipation for at-speed testing of some designs. Note that the same problem occurs in scan-based testing using external ATE (with or without compression). However, ATPG patterns are easier to control~\cite{BuSJ04,LiRP05}. For example, don't care bits of ATPG patterns can be filled in a way which reduces switching activity during the capture cycle~\cite{WeYM05,ReXZ06,WeNM12}. 
ATPG patterns for large designs typically contain 95\%-99\% of don't care bits~\cite{WaC05}.
Since LBIST is based on pseudo-random patterns which are fully specified, they cannot be controlled in the same manner.

Two undesirable consequences of the increased switching activity are overheating and false positives (reporting a fault-free circuit as faulty). The overheating (global or local) decreases the reliability of a circuit, shortens its life time, and may even damage it. False positives can be produced by LBIST due to the non-relevant delay faults caused by IR-drop or crosstalk. 

{\em IR-drop} is the amount of change in the power/ground rail voltage due to the resistance of devices between the rail and a node of interest in the circuit under test~\cite{Sa03}. As process technology scales below 65nm, sensitivity of the speed performance of the circuit to voltage increases. For example, at 65nm and nominal voltage 1V, 5\% voltage change impacts nominal delay by 10.5\%. At 40nm and nominal voltage 0.9V, 5\% voltage change impacts nominal delay by 14.5\%~\cite{Kr08}. This degradation in performance grows worse as the frequency of testing increases. So, a device which passes a slow-speed test at nominal voltage or even $V_{min}$ ($V_{DD}$ - 10\%) conditions might fail a test when the capture frequency is raised.

The underlying mechanism in {\em crosstalk} is related to capacitive coupling between neighbouring nets within a chip~\cite{Sa03}. If the aggressor and victim switch together in the same/opposite direction, the delay of the victim decreases/increases. Depending on the degree of coupling and the switching activity of aggressors/victims during at-speed test, a fault not detectable in the functional mode may occur in the test mode.

While the problem of increased switching activity during shift cycle can be mitigated 
by using a slower clock for shift, to our best knowledge, no good solution for reducing switching 
activity during capture cycle exists at present. In our opinion, this is one of the major 
obstacles for successful deployment of LBIST in the industrial practice.

Another important problem is test coverage of LBIST.
The difficulty of this problem varies from design to design. 
In some cases, LBIST can detect 99\% of transition faults, 
as in the notorious example of IBM S/390 zSeries 900 Microprocessor~\cite{Ku01}.
But there are also cases when LBIST can only detect 65\%-80\% of stuck-at faults~\cite{DaT00}. 

Finally, test execution time of the traditional LBIST may be too long for some applications. 
If the primary reason for using LBIST is in-field testing, then restrictions on test execution time might be very sharp. For example, for Radio Base Stations (RBS), it is typically expected that LBIST reaches the test coverage of 90\% stuck-at faults and 75\% of transition faults within 10 sec. 

Due to the above mentioned problems, today many ASIC vendors are reluctant to use LBIST. Their design flows are typically oriented towards test compression and do not incorporate LBIST. 

\section{What has been done} \label{prev}

In order to address IR-drop, crosstalk and overheating issues,
which become more severe with higher levels of switching activity, 
a number of techniques for reducing power dissipation have been proposed (see~\cite{GoC09} for an excellent overview), including:
\begin{itemize}
\item Designing a better power grid which takes both, functional operation and test mode, into account.
\item Using On-Chip Clock (OCC), which uses full-speed clock for capture and slower clock for shift, as in test compression. 
\item Adding vias to increase the amount of current on timing-critical paths, as in yield improvement.
\item Subdividing the design so that only a part of it is tested at a time.
\item Weighting the pseudo-random patterns so that 0s and 1s occur with a different probability.
\end{itemize}

However, available subdivision methods increase test execution time and they may not lower peak power dissipation sufficiently so that each of the  partitions is operable and at-speed testable, especially under $V_{min}$ condition.
It is our experience that existing weighting techniques cannot reliably control switching activity during capture cycle for 65nm technology and below.

Several methods for increasing test coverage of LBIST have been proposed, including modification of the circuit under test by inserting test points into the circuit~\cite{EiL83,ToM96,TaR96}, modification of the LFSR to generate a weighted sequence with a different distribution of 0s and 1s~\cite{ChM84}, and
embedding of deterministic test patterns into LFSR's patterns by 
LFSR re-seeding~\cite{Ko91} or pattern matching~\cite{GuP88}.

In {\em LFSR reseeding} schemes, deterministic test patterns are encoded into the \emph{seeds} which are loaded as state vectors into an LFSR.
The encoding is done by solving a system of linear equations.
Successful encoding of a pattern into a seed is not guaranteed.
However, as shown by K{\"o}nemann, the probability of encoding failure can be reduced to $1/10^6$ 
by selecting the LFSR of size $S_{\textnormal\small{max}}+20$~\cite{Ko91}.
The encoding efficiency can be increased by using variable-length seeds and multiple polynomials~\cite{HeTRC92,RaTZ98}, or through partial dynamic reseeding technique~\cite{KrJT01}.
The seeds can be stored in an on-chip ROM. 
An alternative approach is to dynamically generate the seeds using a reseeding circuit~\cite{AlM03}.
The order of the seeds might affect the size of the reseeding circuit, or even the number of seeds.
A seed ordering technique which minimizes the area overhead is presented in~\cite{AlMM03} .

In {\em pattern mapping} approaches, a {\em mapping circuit} is placed in between an LFSR and a circuit under test to transform the pseudo-random patterns into deterministic patterns~\cite{Wu98}.
The pattern mapping technique presented in~\cite{ChP95} uses Generalized LFSRs (GLFSRs) as the random pattern generators, and the mapping circuit for each output is synthesized separately.
Another class of pattern mapping techniques includes bit-flipping~\cite{WuK96} and bit-fixing~\cite{ToM96_bitfix} schemes.
They exploit the fact that in a carefully selected random pattern  
only a few bits have to be altered in order to make it deterministic.
In the bit-flipping approach, the mapping circuit implements a Boolean function which evaluates to 1 whenever a flip of a bit is required~\cite{WuK96}.
The bit-fixing logic generates output signals indicating whether a bit should be fixed to 0, to 1, or left unchanged~\cite{ToM96_bitfix}.
A random pattern generated by an LFSR is then modified according to the output of the bit-flipping or bit-fixing function to form the deterministic pattern.

\begin{figure}[t]
\begin{center}
\includegraphics*[width=3.5in]{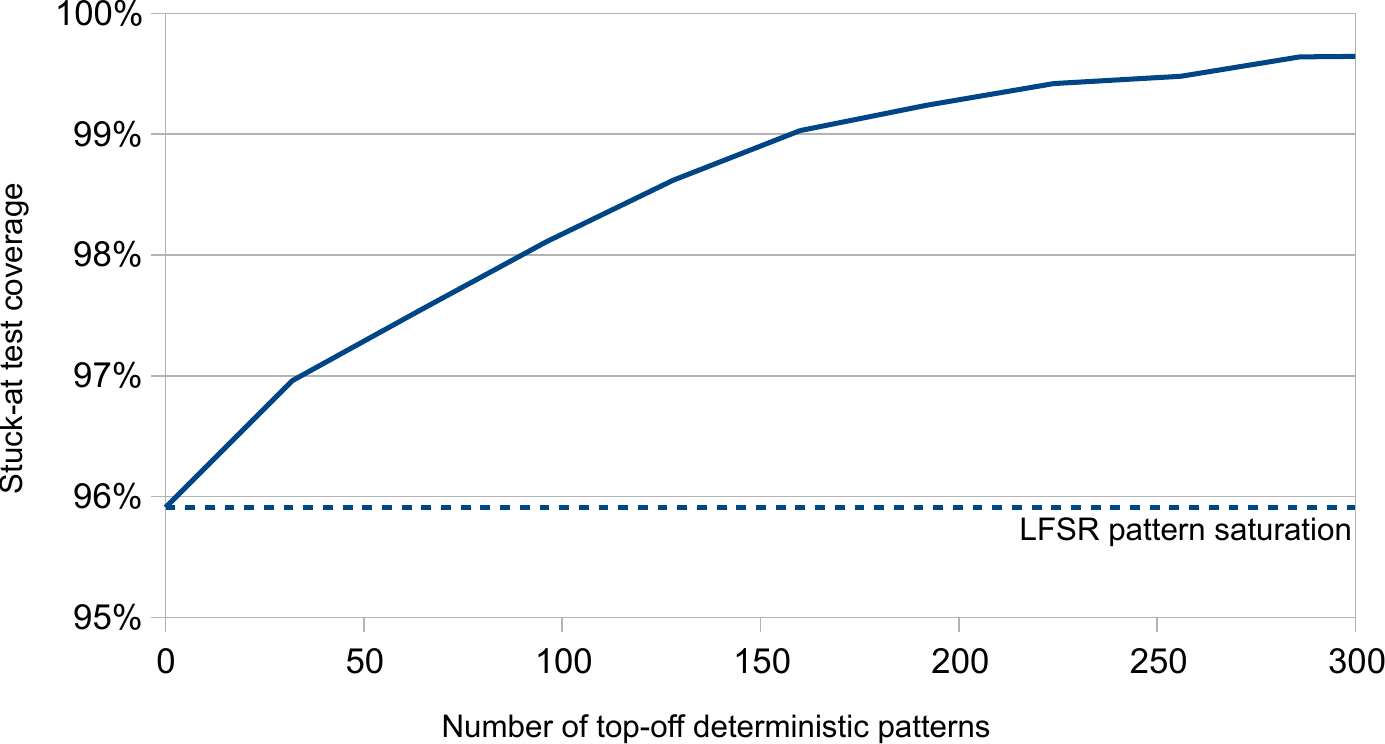}
\caption{Increasing stuck-at test coverage of Leon3 processor using deterministic top-off patterns.}\label{leon_top}
\end{center}
\end{figure}

Yet another approach for increasing test coverage of LBIST is to complement pseudo-random patterns by deterministic top-off patterns~\cite{bist}. Figure~\ref{leon_top} shows an example for Leon3 processor.
We can see that, by using top-off patterns, the stuck-at test coverage of Leon3 can be increased beyond the LFSR pattern saturation point near 96\%. In order to reach 99\% test coverage, which is typically required for production testing, it is sufficient to add 160 deterministic patterns 
on the top of 20.000 LFSR patterns.
Top-off patterns can be stored in an on-chip memory~\cite{SaDB84} or encoded in an finite state machine~\cite{LiD14}. 
This approach is particularly attractive because deterministic top-off patterns can also solve 
the problem with transition or delay faults which are not handled efficiently by the
pseudo-random patterns. 
However, the area required to store top-off deterministic test patterns
within a system can be prohibitively high.
The memory required to store them may exceed 30\% of the memory used in the conventional ATPG-based approach~\cite{HeF99}.

\section{What needs to be done} \label{needs}

We believe that the following problems need to be addressed to successfully deploy LBIST in the industrial practice. 

\begin{enumerate}
\item Subdivision techniques need to be further developed for a variety of design situations. They should require a minimal design modification and should not cause a performance degradation. 
The analysis of switching activity during LBIST versus normal mode can be used to guide subdivision.
By following the hierarchical design-for-test approach and testing only portions of the design at a time, we
might be able to successfully mitigate the problems with false positives and overheating.

\item Alternative ways of weighting pseudo-random patterns to reduce switching activity 
have to be explored. The goal should be to reduce peak and average power dissipation during capture cycle without sacrificing test coverage and test execution time.

\item Techniques for storing deterministic top-off test patterns on-chip with the minimal area overhead are needed. Deterministic top-off test patterns can be viewed as incompletely specified random binary sequences. Better data structures and optimization algorithms for such sequences are required. Compression techniques can be useful in this context.

\item More research on test-per-clock architectures, in which a test pattern is applied to the circuit under test at each clock cycle, is required. Test-per-clock approach considerably reduces the test application time~\cite{SiSS03}. However, it increases the area overhead. An approach achieving a good trade-off between the test application time and area overhead is needed.

\item LBIST methods which take advantage of multiple identical blocks/cores on a chip need to be developed. Existing LBIST CAD tools do not exploit this possibility.
For example, to reduce the area overhead of LBIST, the same pseudo-random test pattern generator can be used for testing identical blocks.

\item It might be more efficient to test regular structures, such as crossbars and switches, using algorithmic test patterns rather than LBIST. Methods for generating such patterns are required.
\end{enumerate}

It would also be advantageous to use LBIST not only for off-line testing, but also for 
predictive maintenance. For example, LBIST can be used for monitoring the condition of back-up modules. This information, together with the data gathered from the operative modules, would help making a decision when system maintenance should be done.
Extending this idea further, we may consider partitioning a system
and scheduling LBIST so that it is applied to inoperative sub-parts of the system or parts 
which operate with reduced functionality. 

It is important to point out that the traditional LBIST techniques target random faults only. 
They do not provide an adequate protection against malicious circuit alternations known as hardware Trojans~\cite{TeK10}. Trojans make possible to bypass or disable the security of a system. The purpose of Trojan insertion can be either to leak confidential information to the adversary, or to disable/destroy a chip. 

There are two different kinds of Trojans~\cite{TeK10}. 
{\em Functional} Trojans add or remove transistors, gates or other components to/from the original design.
{\em Parametric} Trojans reduce the reliability of a chip by thinning of wires, weakening of transistors, or subjecting the chip to radiation. A chip with a parametric Trojan produces errors or fails every time the affected component is loaded intensely.
 
The recent attack on the random number generator of Intel's Ivy Bridge processor~\cite{BeRPB13} demonstrated that the traditional LBIST
may fail even the simple case of stuck-at fault type of Trojans.
This attack was done by modifying the dopant masks to shorten the outputs of selected gates to GND or to $V_{DD}$. The points of modifications were selected so that the compacted signature computed by the MISR for the Trojan-injected circuit coincided with the fault-free circuit signature.  This shows that the traditional LBIST methods need to be strengthened to resist malicious faults as well. 
It is most likely that, in order to be able to perform trusted operations on untrusted hardware, a combination of countermeasures will be required. 
We believe that combining LBIST with the techniques for fault-tolerant design~\cite{Du13book} might be very helpful in this context.
For example, methods for hardening hardware to work in harsh environments can be applied to improve the reliability.

It is likely that, apart from LBIST, new techniques for the mitigation of expected reliability decrease
will be needed. Furthermore, different forms of self-repair of logic to compensate for permanent 
faults that occur during operation will be required. Soft repair technology for memories was introduced by ASIC vendors already at 40/45nm.
Since the geometries of logic are less dense, and hence 
more robust, we expect the error rates of logic to be similar to those of memories one or two process 
generation later.

\section{Conclusion} \label{con}
 
The goal of this paper is to attract a wider attention of research community to the technical 
problems which prevent a successful deployment of LBIST in the industrial practice.
We discussed what needs to be done and gave some thoughts on how these issues can be addressed.
We look forward to a continued dialogue in this area.


\balance
\bibliographystyle{ieeetr}
\bibliography{bib}





\end{document}